\documentclass[preprint,aip]{revtex4-1}
\usepackage{amsmath}
\usepackage{units}
\usepackage{graphicx}
\usepackage{comment}
\usepackage{float}
\usepackage{epstopdf}
\usepackage{silence}
\begin{document}

\title{Why are Fluid Densities So Low in Carbon Nanotubes?}

\author{Gerald J. Wang and Nicolas G. Hadjiconstantinou}
\affiliation{Department of Mechanical Engineering, Massachusetts Institute of Technology, Cambridge MA}
\date{\today}
\begin{abstract}
The equilibrium density of fluids under nanoconfinement can differ substantially from their bulk density. Using a mean-field approach to describe the energetic landscape near the carbon nanotube (CNT) wall, we obtain analytical results describing the lengthscales associated with the layering observed at the interface of a Lennard-Jones fluid and a CNT. We also show that this approach can be extended to describe the multiple-ring structure observed in larger CNTs. When combined with molecular simulation results for the fluid density in the first two rings, this approach allows us to derive a closed-form prediction for the overall equilibrium fluid density as a function of CNT radius that is in excellent agreement with molecular dynamics simulations. We also show how aspects of this theory can be extended to describe some features of water confinement within CNTs and find good agreement with results from the literature.
\end{abstract}
\maketitle

\section{Introduction}
Fluids under nanoscale confinement exhibit many remarkable properties \cite{Hummer, Majumder, Holt, Whitby}. Of particular interest is the observation that when a carbon nanotube (CNT) is in equilibrium with a fluid bath, the density of the fluid inside the CNT can differ dramatically from the density of the bulk fluid -- this value can be as low as $200$ kg m$^{-3}$ for nanoconfined water \cite{Kassinos}. Understanding and predicting this anomaly is very important for a variety of applications, such as designing nanoscale desalination devices \cite{Hinds}, engineering nano-syringes for drug delivery across cell membranes \cite{Nanosyringes}, calculating shale gas or oil content of nanoporous rock \cite{Bernard}, and potentially for assisting with the development of models that predict anomalous fluid flow rates through CNTs\cite{Grammenos, Takaba, TruskettWater}. Predicting equilibrium densities under confinement can also be very beneficial from a computational point of view, because it allows realistic simulation of nanofluidic systems without coupling to an external fluid bath \cite{Wu, Das, Gordillo}. Benefits are possible even when a fluid bath is included in such systems: for example, it is common to pre-fill nanopores with fluid molecules to reduce equilibration time; knowledge of the {\it correct} equilibrium density minimizes the computational cost associated with equilibration.

\begin{figure}[H]
\begin{centering}
\includegraphics[width=0.45\textwidth]{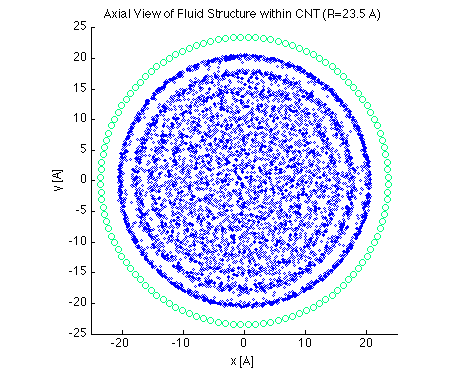}
\caption{\label{sample_rings} Cross-sectional view of equilibrium LJ fluid structure within a CNT ($R=23.5\AA$), obtained from MD simulation. Green circles (lighter gray) denote wall carbon atoms and blue dots (darker gray) denote fluid atoms.}
\end{centering}
\end{figure}

Several molecular dynamics (MD) studies have investigated these anomalous equilibrium densities under nanoconfinement \cite{Travis, Kassinos, Wu, JunWang, LiuWang} but there is currently no first-principles model that can predict this density without the high computational costs of a density-functional theory calculation \cite{Peng}. These studies have established that fluids confined within a sufficiently large CNT will form concentric rings near the CNT wall \cite{Kassinos, Wu, JunWang, LiuWang}. Near the center of the CNT, the fluid will exhibit little ordering and resemble bulk fluid, or fluid that is not ``aware" of the presence of the CNT wall. These features can be seen in Fig. \ref{sample_rings}, which shows the structure of a Lennard-Jones (LJ) fluid in a CNT of radius 23.5$\AA$. Numerous studies \cite{Hanasaki, Nozzles, ThomasMcGaughey, Travis} have observed the presence of a stand-off distance between the CNT wall and the fluid, determined empirically to be on the order of one atomic diameter. 

In this paper, we present a classical mean-field approach that provides an accurate analytical prediction for the stand-off distance between the fluid and the CNT wall. Comparison with MD simulation results shows that the prediction for the width of this excluded volume region is very accurate for $R>5\AA$, but remains reliable even for $3\AA\lesssim R\lesssim 5\AA$, where single-file flow is observed (for $R\lesssim 3\AA$ imbibition is not possible \cite{Kassinos}). We also show that this approach can be extended to the calculation of the lengthscales associated with the multiple-ring structure observed in larger ($R\gtrsim 9\AA$) CNTs -- that is, to predict ring locations, thicknesses, and (consequently) the excluded volumes between rings. We finally couple this description with MD results for the density inside the first two rings to derive an expression for predicting the equilibrium density of a LJ fluid inside ``large" ($R\gtrsim 9\AA$) CNTs as a function of the CNT radius. We compare our results to MD simulations -- both in-house and from other research groups -- and find excellent agreement. We also show that certain aspects of this theory (namely, the maximum radius accessible to the fluid) can be extended to water confined within CNTs; again, excellent agreement with MD simulations is observed.

\section{Mean-field description}
In this work we focus on intermolecular interactions governed by the Lennard-Jones potential \cite{Allen} 
\begin{equation}\label{LJ}
V(r)=4\varepsilon \bigg[\Big(\frac{\sigma}{r}\Big)^{12}-\Big(\frac{\sigma}{r}\Big)^6\bigg]
\end{equation}
where parameters $\varepsilon$ and $\sigma$ denote interactions between a fluid molecule and a wall carbon; interactions between fluid molecules will be denoted by $\varepsilon_f$ and $\sigma_f$. 

Assuming that the CNT is sufficiently long compared to its radius $R$ and that the radius is sufficiently large so that the CNT can be approximated as cylindrical, we can derive a mean-field interaction potential between the CNT wall and the fluid by integrating the LJ potential around the cylindrical geometry of the CNT \cite{TFM} to obtain
\begin{eqnarray} \label{LJ_cyl}
\mathcal V(r)&=&n\pi^2\varepsilon\sigma^2\bigg[\frac{63}{32}F_{-\frac{9}{2};-\frac{9}{2};1}(\delta^2)\bigg(\frac{R(1-\delta^2)}{\sigma}\bigg)^{-10}-3F_{-\frac{3}{2};-\frac{3}{2};1}(\delta^2)\bigg(\frac{R(1-\delta^2)}{\sigma}\bigg)^{-4}\bigg]
\end{eqnarray}

Here, $n$ denotes the areal density of carbon atoms in the CNT wall, $\delta$ the normalized radius $\delta\equiv r/R$, and $F_{\alpha;\beta;\gamma}(z)$ the Gauss hypergeometric function \cite{Abramowitz}. In what follows, $F_{\eta}(z)$  will be used to denote  $F_{\eta;\eta;1}(z)$. We note that our results in the next section will show that for $R\gtrsim5\AA$, the approximation of the CNT by a circular shape introduces very little error; therefore, given that CNT imbibition is only possible for $R\gtrsim 3\AA$, this approximation is not very restrictive\cite{Kassinos}. 

\begin{figure}
\includegraphics[width=0.7\textwidth]{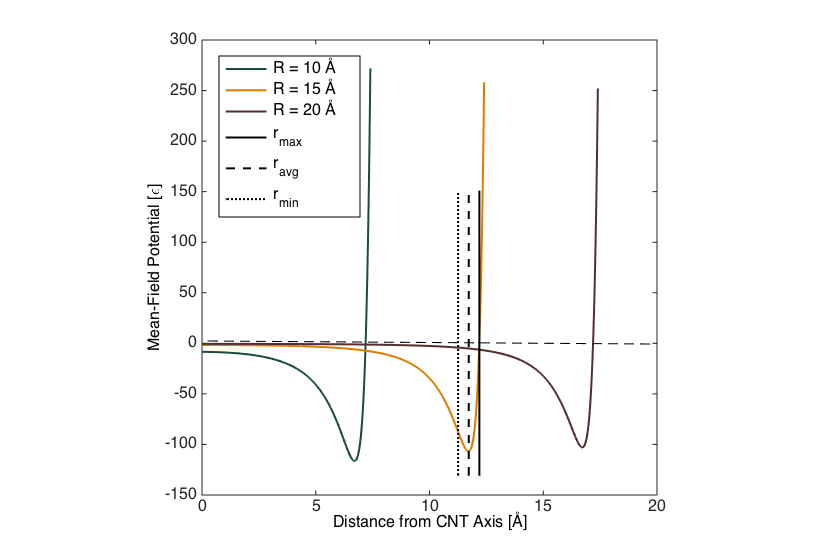}
\caption{\label{mean_field} Mean-field potentials due to CNTs of three different radii (from left to right, $R=10\AA$, $R=15\AA$, $R=20\AA$); $r_\text{max}$, $r_\text{avg}$ and $r_\text{min}$  are labeled for CNT with $R=15\AA$.}
\end{figure}

The mean-field potential for a variety of CNT radii is shown in Fig. \ref{mean_field}. As expected, it rises sharply as $r\rightarrow R$, thus leading to a maximum radius that is energetically accessible to the fluid. This radius will be referred to as $r_\text{max}$. Our work below exploits this very steep rise in the mean-field potential to obtain an analytical result for $r_\text{max}$. In larger CNTs, where multiple rings form (as is the case in Fig. \ref{sample_rings}), $r_\text{max}$ will represent the outer radius of the first ring. A methodology for calculating the first ring thickness is given in Section \ref{sectionB}, while the structure of subsequent rings and excluded volumes between rings are discussed in Section \ref{sectionC}.

\subsection{Maximum accessible radius}
We determine the maximum accessible radius $r_\text{max}(R,\sigma, \varepsilon, T)$ by finding the location at which most of the fluid molecules have insufficient kinetic energy to overcome the potential barrier $\mathcal V (r)$. The steep rise of $\mathcal V(r)$ close to $r=R$ (where $r_\text{max}$ is expected to lie) allows us to approximate this location by $\mathcal V(r_\text{max})=0$ with little error for typical LJ parameters $\varepsilon$ and $\sigma$ and temperature $T$. A direct consequence is that our solution for $r_\text{max}$ is independent of the temperature.

Setting $\mathcal V(r_\text{max})=0$, we can rearrange \eqref{LJ_cyl} into
\begin{equation}\label{LJ_cyl_rearr}
\frac{21}{32}\sigma^6=R^6\Big(1-\delta_\text{max}^2\Big)^{6}\frac{F_{-\frac{3}{2}}\big(\delta_\text{max}^2\big)}{F_{-\frac{9}{2}}\big(\delta_\text{max}^2\big)}
\end{equation}
where $\delta_\text{max}=r_\text{max}/R$.
This equation shows that the parameter $\varepsilon$ can be scaled out of the problem and thus $\delta_\text{max}=\delta_\text{max}(R,\sigma)$.
To solve, we write $\delta_\text{max}=1-k_\text{max}\sigma/R$, where $k_\text{max}\sigma$ is the stand-off distance from the CNT wall and $k_\text{max}=k_\text{max}(R,\sigma)$. This choice is motivated by the expectation that the standoff distance will be of order $\sigma$. 
Inserting the above expression for $\delta_\text{max}$  in \eqref{LJ_cyl_rearr} we obtain to leading order 
$k_\text{max}(R,\sigma)= (2/5)^{1/6}$,
which yields
\begin{equation}\label{MAR}
r_\text{max}(R,\sigma)=R-(2/5)^{1/6}\sigma
\end{equation}
The above result was obtained by using a theorem due to Gauss \cite{Abramowitz} to expand the hypergeometric function near $\delta_\text{max}=1$ in the form
\begin{equation}
F_{\alpha;\beta;\gamma}(\delta_\text{max}^2)= \frac{\Gamma(\gamma)\Gamma(\gamma-\alpha-\beta)}{\Gamma(\gamma-\alpha)\Gamma(\gamma-\beta)} \Bigg(1+\mathcal O \bigg(\frac{\sigma}{R}\bigg)\Bigg)
\end{equation}

To validate the analytical result \eqref{MAR}, we solved equation (\ref{LJ_cyl_rearr}) numerically and conducted MD simulations (methodology described in Appendix); the results are shown in Figure \ref{standoffvssigma}. These figures show that \eqref{MAR} is in agreement with but also explains our MD results as well as MD results by other groups \cite{ThomasMcGaughey, Hanasaki} for $R\gtrsim 5\AA$. 

In agreement with our model prediction, our MD simulations (in the temperature range 100K$\leq T \leq$ 400K) as well as simulations from other groups \cite{TempEffects, HungGubbins, Zhao} show negligible dependence on temperature.  We also note that the result \eqref{MAR} is valid for a wide range of bulk fluid densities; specifically, our MD simulations covered the range $0.8\sigma_f^{-3}\leq\rho_\text{bulk}\leq 1.1\sigma_f^{-3}$. The bulk density, $\rho_\text{bulk}$ is defined as the density of the bulk fluid with which the fluid in the CNT is in equilibrium; in our simulations it was imposed by placing the CNT in a finite but large reservoir of fluid. More details can be found in the Appendix.

The leading-order solution obtained here is equivalent to neglecting the effect of CNT curvature, explaining why the stand-off distance is not a function of $R$. Although inclusion of higher-order terms in the solution is possible, the excellent agreement of \eqref{MAR} with numerical solution of \eqref{LJ_cyl_rearr} as well as with MD simulations suggests that higher-order terms are unnecessary. 

We also note that $R\approx 5\AA$ is approximately equal to the largest radius at which single-file imbibition is observed \cite{Kassinos}. In other words, the above result is valid even before a ring structure is visible; in the more general case, $r_\text{max}$ can be  identified with the location at which the fluid radial density function (RDF) vanishes.
\subsection{First ring thickness and inner radius}
\label{sectionB}
In this section we consider CNTs that are sufficiently large ($R\gtrsim 6\AA$) that at least one fluid ring has clearly formed within the CNT cross-section. In this case, $r_\text{max}$ will correspond to the outer radius of this ring. To describe the thickness of the ring, we also need the ring inner radius. This quantity can be calculated by again exploiting knowledge of the shape of $\mathcal V(r)$ as $r\rightarrow R$. Specifically, we assume that the first ring is centered in a symmetric fashion around the minimum of $\mathcal V(r)$, denoted by $r_\text{avg}$. We estimate the half width of this ring as $r_\text{max}-r_\text{avg}$, and so the location of the inner radius can be calculated as $r_\text{min}=2r_\text{avg}-r_\text{max}$.

The value of $r_\text{avg}$ is obtained by setting the derivative of \eqref{LJ_cyl} to zero and is given by
\begin{eqnarray}\label{ravg_mess}
\frac{\delta_\text{avg}}{R^{11}}\bigg[\frac{81F_{-\frac{7}{2}}(\delta_\text{avg}^2)}{2(1-\delta_\text{avg}^2)^{10}}+ \frac{20 F_{-\frac{9}{2}}(\delta_\text{avg}^2)}{(1-\delta_\text{avg}^2)^{11}}\bigg]- \frac{32\delta_\text{avg}}{21 \sigma^6R^5}\bigg[\frac{9F_{-\frac{1}{2}}(\delta_\text{avg}^2)}{2(1-\delta_\text{avg}^2)^{4}}+ \frac{8 F_{-\frac{3}{2}}(\delta_\text{avg}^2)}{(1-\delta_\text{avg}^2)^{5}}\bigg]=0
\end{eqnarray}
where $\delta_\text{avg}=r_\text{avg}/R$.

Following the same argument as in the previous section, we write $\delta_\text{avg}$ in the form $\delta_\text{avg}(R,\sigma)=1-k_\text{avg}\sigma/R$. We proceed to solve \eqref{ravg_mess} by neglecting terms smaller than $\mathcal O\big(R/\sigma\big)$ to obtain
\begin{equation}
\frac{21}{32}\sigma^6 =\frac{\frac{256}{3\pi}R\Big(2k_\text{avg}\sigma\Big)^{-1}}{\frac{524288}{63\pi}R\Big(2k_\text{avg}\sigma\Big)^{-7}}
\end{equation}
which simplifies to $k_\text{avg}= 1$.
This means that the midpoint of the outer ring is at a distance $\sigma$ from the CNT wall. Therefore, $r_\text{min}=R-k_\text{min}\sigma$, where $k_\text{min}=2-(2/5)^{1/6}$; $k_\text{min}\sigma$ represents the distance between the inner radius of the ring and the CNT wall. 

In the following section we show how this methodology can be extended to the description of subsequent rings that appear as the CNT radius increases.

\subsection{Subsequent rings}
\label{sectionC}
In this section, we consider CNTs for which at least two distinct rings appear before relaxation to bulk structure occurs ($R\gtrsim 9\AA$). We capture the geometry of additional rings within the outermost ring by recognizing that the outermost ring itself can be treated as another CNT. This general approach of recognizing that a cylindrical solid structure induces concentric near-solid ordering in adjacent fluid has been pursued with success by Wilson in numerous MD studies \cite{Wilson2, Wilson1}. 

\begin{figure}
\includegraphics[width=0.55\textwidth]{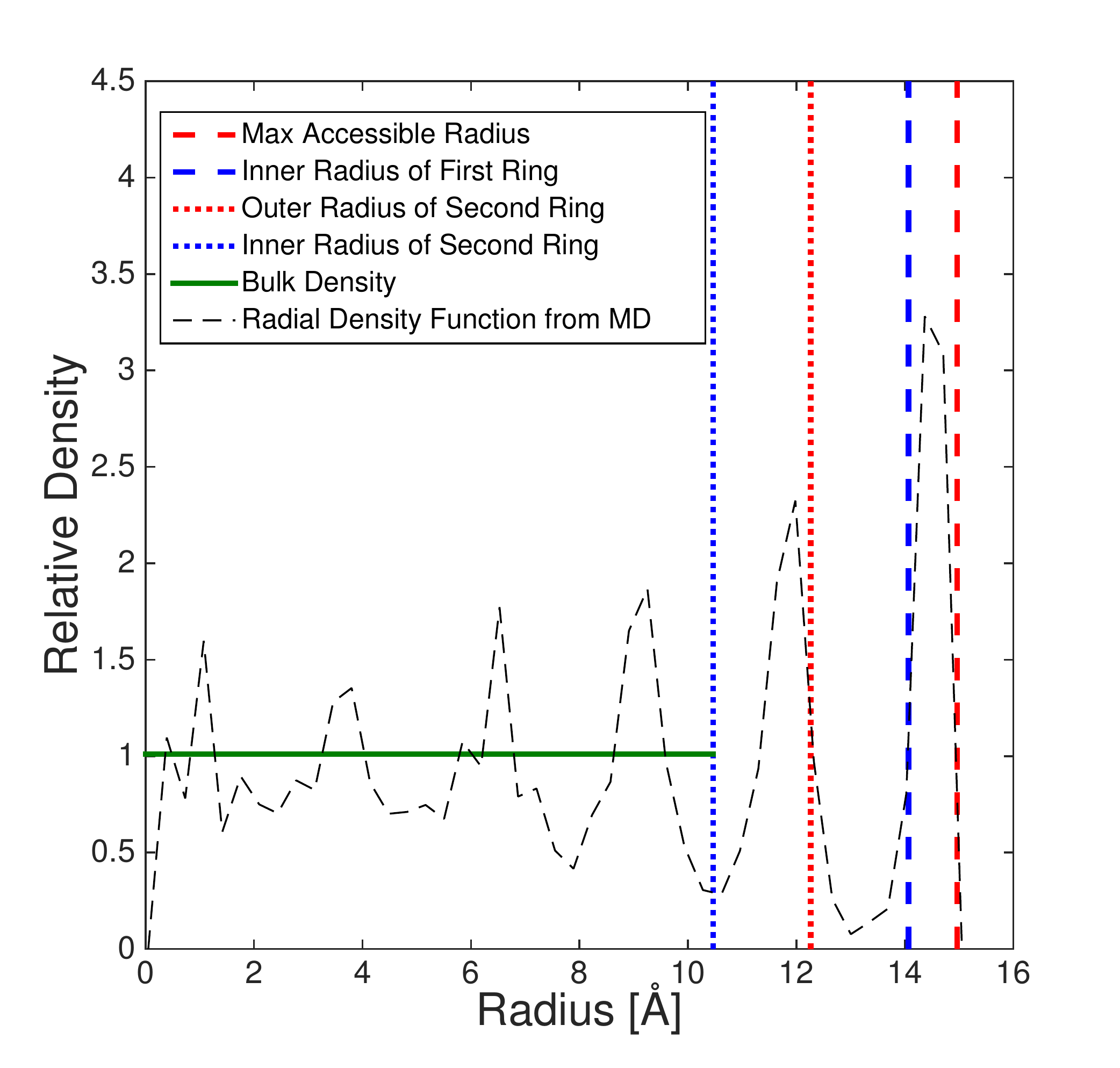}
\caption{\label{RDF_3945} Theoretical prediction for ring locations in a CNT (R = 17.61 \AA) and radial density profile from MD simulation.}
\end{figure}

Let us denote the outermost ring's outer radius by $r_\text{(1),max}$ and its inner radius by $r_\text{(1),min}$. Then an estimate for the second ring's outer radius is $r_\text{(2),max}=r_\text{(1),max}-k_\text{max}\sigma_f$ and a bound for its inner radius is $r_\text{(2),min}=r_\text{(1),min}-k_\text{min}\sigma_f$. In principle, this bounding process can be repeated indefinitely to fix outer and inner radii for the $n$-th ring $r_{(n)}$, but in practice this method loses meaning after the outer radius of the $(j+1)$-st ring is greater than the inner radius of the $j$-th ring, at which point the rings are ``blurred" into a more uniform background bulk structure. For the purposes of this study, we will only consider two rings (in addition to the bulk core). Fig. \ref{RDF_3945} shows a comparison between these analytical expressions and our MD simulation results for a temperature of 300K and $\rho_\text{bulk}=1 \sigma_f^{-3}$. The good agreement extends to the whole range of simulations performed in this work ($0.8\sigma_f^{-3}\leq \rho_\text{bulk}\leq 1.1 \sigma_f^{-3}$ and 100K $\leq T \leq$ 400K).

\begin{figure}
\includegraphics[width=0.55\textwidth]{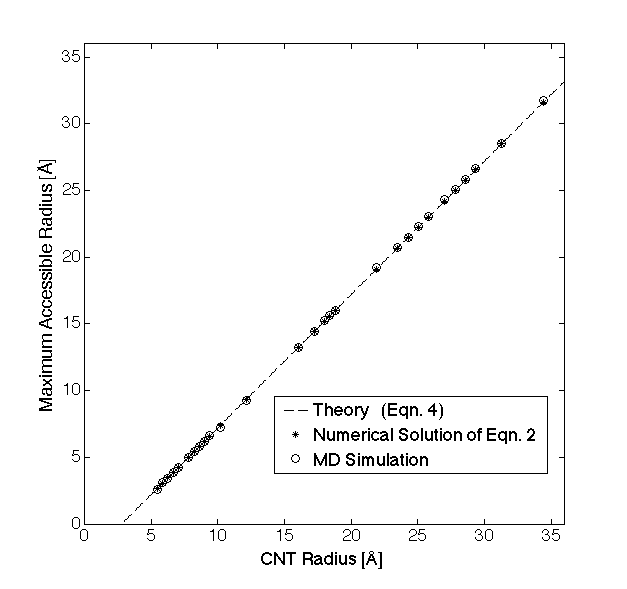}
\caption{\label{standoffvssigma} Comparison between theoretical prediction for maximum accessible radius \eqref{MAR}, numerical solution of \eqref{LJ_cyl_rearr}, and maximum accessible radius from MD simulations.}
\end{figure}

\section{Predicting Fluid Density in ``Large" CNTs}
We now show that combining the above predictions for the ring locations and widths with information from MD simulations about the fluid density inside the first two rings yields a closed-form expression for the equilibrium fluid density inside a CNT as a function of its radius. This formulation is intended for CNTs with $R\gtrsim 9\AA$ i.e. radii that are sufficiently large for the ``fully developed" ring structure described in Section \ref{sectionC} (two distinct rings and a bulk core) to exist. 

The fluid density in ring $i$ is defined as the number of fluid-molecular centers falling within the range $[r_\text{(i),min},r_\text{(i),max}]$, divided by the volume enclosed by this region. We note that this simple definition captures over 93\% of molecules in the ring region; the small number of remaining molecules are assigned to the nearest ring. The mean ring densities obtained from MD simulations over a range of CNT radii are reported in Table \ref{density_table}. We note that the relatively high densities reported in this table, especially for the outermost ring, are a result of using the volume which encloses all atomic centers to define density. If, for example, one extends this volume by $0.5\sigma$ in each direction to account for the true volume occupied by the atoms, the ring densities would be much closer to unity.

\begin{table}[h]
\begin{center}
\begin{tabular}{c | c}
Ring & Normalized Density\\
\hline
1st & $\rho_{(1)}=3.15 \pm 0.04$ \\
2nd & $\rho_{(2)}=1.40 \pm 0.07$ \\
Bulk core & $1.01 \pm 0.02$ \\
\end{tabular}
\end{center}
\caption{Densities of each ring, calculated from MD simulation at $T=300K$, normalized by the bulk density.}
\label{density_table}
\end{table}

\label{overall_density}
Combining the predicted locations and widths of the rings with the ring densities calculated from MD simulation, we can construct the following expression for the overall normalized density as a function of CNT radius $R$:
\begin{eqnarray}\label{densityexpr}
\rho(R)=\frac{1}{R^2}\Big((r^2_\text{(1),max}-r^2_\text{(1),min})\rho_{(1)}+(r^2_\text{(2),max}-r^2_\text{(2),min})\rho_{(2)}+r^2_\text{(3),max}\Big)
\end{eqnarray}

It can be readily verified that this expression asymptotically approaches unity for large $R$, as expected. Here it is important to recall that this density is normalized by the bulk density $\rho_\text{bulk}$ of the fluid with which the fluid in the CNT is in equilibrium; in other words, for a CNT placed in a bath of fluid at density $\rho_\text{bulk}$, the density of the fluid in a CNT of radius $R$ is $\rho(R) \rho_\text{bulk}$.  

\begin{figure}
\includegraphics[width=0.55\textwidth]{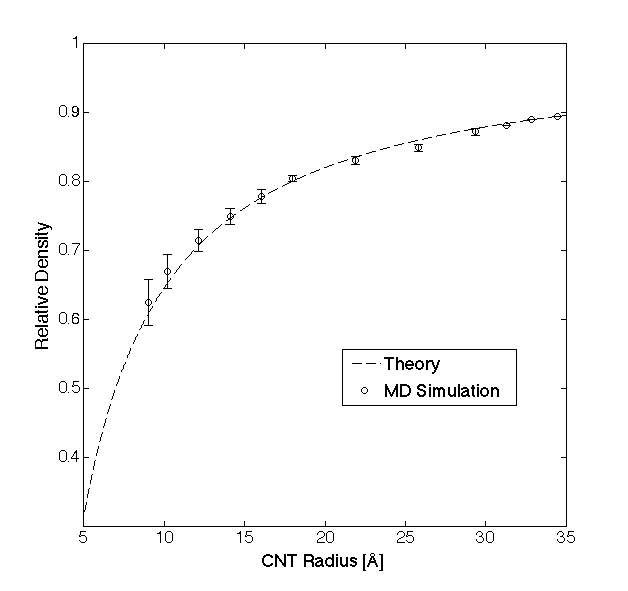}
\caption{\label{expvstheory} Theoretical prediction for $\rho(R)$ with densities measured from MD simulations overlaid.}
\end{figure}

Figure \ref{expvstheory} shows that \eqref{densityexpr} is in excellent agreement with actual densities measured in MD simulations at a bulk density of $1.0 \sigma_f^{-3}$ and $T=300$K. For $R>15\AA$, the discrepancy is within $3\%$ for all simulated CNTs in the range $0.8\sigma_f^{-3}\leq\rho_\text{bulk}\leq 1.1\sigma_f^{-3}$. 

Although information from MD simulations is still required to determine some of the parameters in (\ref{densityexpr}), we expect this equation to be preferable to the empirical fits proposed previously (for example, in [\onlinecite{Kassinos}]) for a number of reasons. First, it clearly illustrates the physical considerations that determine the average density in a CNT (namely, a volume-weighted average of the density in the rings, the bulk core, and the excluded volumes that appear in this geometry). Second, by including the analytical results about the lengthscales associated with the ring structure, it relies on only two quantities -- with very well defined physical meaning -- that need to be determined from MD simulations.

We close by noting that the density inside the rings is, in general, dependent on temperature. Therefore, given that the results reported in Table \ref{density_table} were obtained for a temperature of 300K, $\rho_{(1)}$ and $\rho_{(2)}$ will need to be calculated from MD simulations or via other means\cite{TruskettAvailableSpace,TruskettStatGeo} if a significantly different temperature is of interest. To test the effect of varying $\varepsilon$, simulations were conducted with $\varepsilon\in\{0.05,0.10,0.25,0.50\}\text{ kJ mol}^{-1}$; it was found that variations over this decade of $\varepsilon$ resulted in changes in the mean density of each ring that are less than 6\% (as compared to results for the baseline value of $\varepsilon=0.48$ kJ mol$^{-1}$). In other words, provided the carbon-fluid interaction is within this hydrophobic range, $\rho_{(1)}$ and $\rho_{(2)}$ are approximately independent of $\varepsilon$. We also recall that the excluded-region lengthscales remain consistent over the full range of simulated conditions described in sections \ref{sectionB}, \ref{sectionC}, and the Appendix.  

\section{Extensions to Nanoconfined Water}
\label{water}
\begin{figure}
\includegraphics[width=0.55\textwidth]{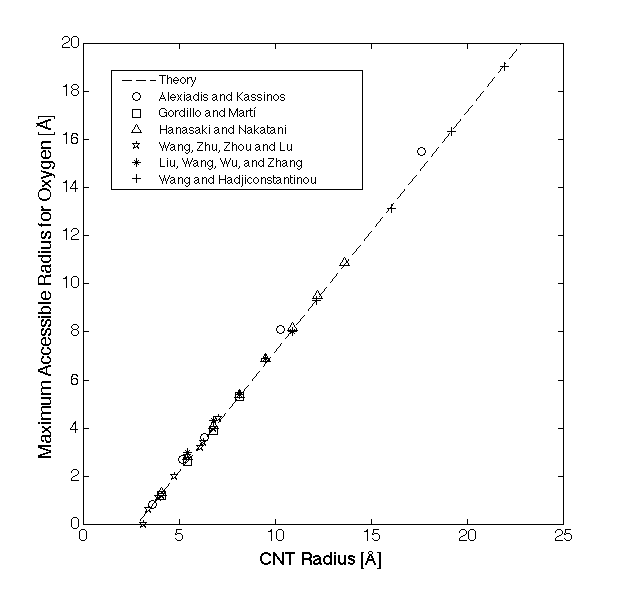}
\caption{\label{MAR_water} Theoretical prediction for maximum accessible radius of the oxygen atom in a water molecule and maximum accessible radius from MD simulations by Refs. \onlinecite{Kassinos}, \onlinecite{Gordillo}, \onlinecite{JunWang}, \onlinecite{LiuWang}, \onlinecite{Hanasaki}, and the authors.}
\end{figure}

In the most common water models for MD simulations (e.g. SPC/E \cite{SPCE}, TIP3P \cite{Jorgensen}, TIP4P \cite{TIP4P}), the interaction between the oxygen atom and other atoms is described by a LJ potential. Since these models assume that LJ interactions between hydrogen and carbon are negligible compared to LJ interactions between oxygen and carbon, we can use \eqref{MAR}  to predict the point at which the oxygen radial density function (RDF) for water confined within a CNT vanishes. These predictions agree closely with results from six sets of MD simulations as shown in Fig. \ref{MAR_water}. The analytical prediction \eqref{MAR} is a definite improvement over current approaches found in the literature, which rely on empirical measurements from MD simulations \cite{Nozzles, ThomasMcGaughey}. It is noteworthy that the presence of electrostatic interactions has a remarkably small effect on the prediction of the zero of the oxygen RDF; this is in agreement with the physical basis of our model, which attributes ring formation to the interplay between short-range attractive and repulsive molecular interactions.

\begin{figure}
\includegraphics[width=0.55\textwidth]{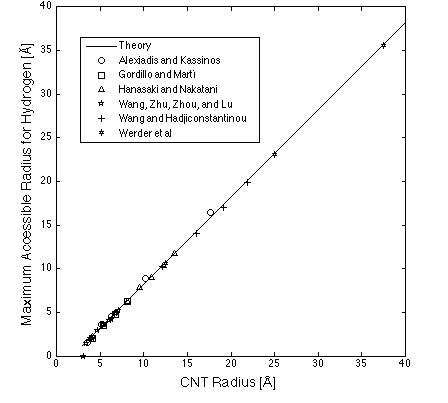}
\caption{\label{MAR_hydrogen} Theoretical prediction for maximum accessible radius of the hydrogen atom in a water molecule and maximum accessible radius from MD simulations by Refs. \onlinecite{Kassinos}, \onlinecite{Gordillo}, \onlinecite{JunWang}, \onlinecite{Hanasaki}, \onlinecite{Werder}, and the authors.}
\end{figure}

In fact, this approach can be used to calculate the maximum accessible radius of hydrogen atoms in water. Since there is no significant interaction between carbon and hydrogen in these water models, the maximum accessible radius of hydrogen is identical to that of oxygen extended by 0.96$\AA$, the size of the O-H bond. Fig. \ref{MAR_hydrogen} shows that this purely geometric argument captures simulation results accurately. 
On the other hand, this geometric argument contains no information about the angular distribution of water molecules \cite{ThomasMcGaughey}. As a result, additional ingredients are required before the equilibrium density of water in CNTs can be described analytically with high precision. This will be the subject of future work.

Despite the above, knowledge of the RDF structure can still be very useful, especially since non-equilibrium MD simulations have shown that fluid flow does not appreciably modify oxygen and hydrogen RDFs from their equilibrium shapes \cite{Hanasaki}. Specifically, this information can be useful in situations where knowledge of the density distribution is useful (e.g. for constructing models of other thermodynamic properties\cite{TruskettThermo, TruskettDumbbell, TruskettWater, TruskettHydrocarbon}), as well as in cases where fluid flow rates are related to ``geometric" effects associated with the volume accessible to the fluid.

\begin{figure}
\includegraphics[width=0.55\textwidth]{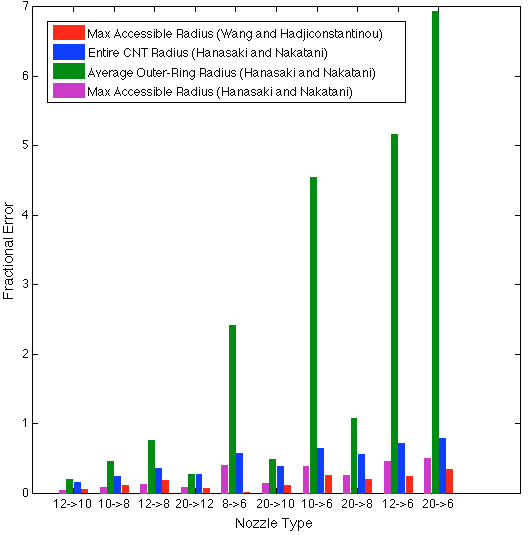}
\caption{\label{nozzles} Fractional error in predictions of velocity enhancement, defined as $(\chi_\text{MD}-\chi_\text{theory})/\chi_\text{MD}$ where $\chi$ refers to the velocity enhancement $\frac{v_\text{downstream}}{v_\text{upstream}}$, using \eqref{MAR} vs. the methods described in Ref. \onlinecite{Nozzles}. A nozzle type of the form $m\rightarrow n$ indicates a constriction from an $(m,m)$ CNT to an $(n,n)$ CNT.}
\end{figure}

As an example of such a geometric effect, we consider the system of CNT junctions described in Ref. \onlinecite{Nozzles}, emulating convergent nozzles. Using MD simulations, Hanasaki and Nakatani found that area ratios based on nominal CNT radii are inaccurate for predicting velocity enhancement downstream of a constriction. It is only when the excluded volume (due to the stand-off distance between the water and the CNT) is taken into account that the simulation results can be reproduced. Fig. \ref{nozzles} shows that using \eqref{MAR} as the basis for a more representative cross-sectional area $\pi r_\text{max}^2$, we find much closer agreement with Hanasaki and Nakatani's MD results. Although Hanasaki and Nakatani ultimately also used a maximum accessible radius, their values for this quantity were only determined empirically.

\section{Conclusions}
Fluid densities are low in CNTs (compared to the bulk density) due to the finite range of molecular interactions, which is non-negligible compared to typical CNT radii. The largest contributor to this phenomenon is the excluded volume between the CNT wall and the fluid, whose thickness is on the order of one molecular diameter. As the CNT radius approaches this lengthscale, the area that contains no fluid molecules becomes a large fraction of the total CNT cross-sectional area. For example, for a CNT with $R\approx 10\AA$, the excluded volume between the wall and the first ring alone accounts for approximately half of the CNT volume. No physically attainable packing within the first ring could yield a sufficiently high first-ring density to make up for this  deficit -- thus, the density of the nanoconfined fluid is quite low relative to the bulk fluid.

To describe this phenomenon, we have developed an energetics approach for predicting the structure of Lennard-Jones fluids confined in CNTs. Although the present study has focused on confinement in CNTs, the basic principle should be applicable to confinement in other geometries as well as other interaction potentials describing simple fluids. We find that electrostatic interactions have a negligible effect in the case of water, which suggests that describing equilibrium densities of complex fluids using this approach may also be possible.

Combining the analytical description of the fluid structure with a characterization of the density in the first two rings from MD simulation provides a closed-form expression for the normalized density of a confined fluid as a function of CNT radius. This expression is in excellent agreement with MD simulations. 

Despite the focus on equilibrium densities, the results discussed here are in some cases already useful for non-equilibrium settings involving fluid flow. For example, our results can be used to estimate the effective cross-sectional area of a CNT, which is vital for accurately predicting the flow velocity enhancement downstream of a constriction. We also note that non-equilibrium MD simulations have shown that water flow in CNTs does not appreciably affect water structure \cite{Hanasaki}. Thus information about the equilibrium spatial distribution of density may also prove useful in the development of models that predict water flow rates in CNTs\cite{Takaba}, a topic which has attracted considerable attention \cite{Hummer, Majumder, Holt, Whitby}.

\section{Acknowledgements}
The authors would like to thank P. Poesio, D. Blankschtein, and R. Kurchin for helpful discussions. This work was supported by Aramco Services and the DOE CSGF under grant number DE-FG02-97ER25308.

\section{Appendix: Molecular Dynamics Simulations} 
We simulated CNTs of radii ranging from $9\AA$ to $35\AA$. All CNTs were of armchair chirality; it has been shown that chirality has no observable impact on equilibrium fluid density within a CNT \cite{Kassinos, JunWang}. For the purpose of calculating density, the CNT radius $R$ is defined as $3an/\pi$, where $(n,n)$ is the chiral vector of the armchair CNT and $a=1.421\AA$, the inter-carbon spacing. Simulations were conducted at 300 K in LAMMPS \cite{LAMMPS} in the NVT ensemble coupled to a Berendsen thermostat \cite{Berendsen}, for reservoir densities $\rho_\text{bulk}\in\{0.8\sigma_f^{-3},1.0\sigma_f^{-3},1.1\sigma_f^{-3}\}$. Additional simulations were performed at 100K and 400K to ensure that the excluded-region lengthscales have no dependence on temperature. The CNT was kept rigid throughout each simulation; it has been shown that thermally oscillating CNT walls have only a small impact on equilibrium fluid structure and density \cite{Zhao}. The simulation time step was 2.0 fs. Each system was allowed to equilibrate for 2.0 ns, after which kinematics were recorded every 2.0 fs for a total of 5.0 ns. To facilitate convergence to equilibrium density, CNTs were pre-filled to a relative density of 0.2. 

To reduce end effects, the length of each CNT was between 2.8 and 3.1 times its diameter. Each CNT was placed in a fluid bath (reservoir) with periodic boundary conditions. The size of the reservoir was at least twice the size of the CNT in each dimension to ensure that the finite bath size did not affect simulation results. The reservoir volume was $100\AA\times100\AA\times120\AA$ for CNTs with $R<20\AA$ and $100\AA\times100\AA\times200\AA$ for CNTs with $R>20\AA$ and was chosen such that the density change in the reservoir due to CNT imbibition was negligible. 

Interactions between carbon and fluid molecules were modeled using $\sigma = 3.28 \AA$ and $\varepsilon=0.48$ kJ mol$^{-1}$ and interactions between fluid molecules were modeled using $\sigma_f = 3.15 \AA$ and $\varepsilon_f=0.64$ kJ mol$^{-1}$. These parameters correspond to interactions between carbon and monatomic oxygen \cite{Jorgensen} and were chosen with simulations of water in CNTs in mind. A cut-off distance of $3\sigma$ was used throughout. Simulations of water utilized the TIP4P potential \cite{TIP4P} with the same LJ parameters governing oxygen-carbon interactions. 

\bibliography{Anomalous_Density_preprint_twocol_V2.bib}

\end{document}